\documentclass[a4paper]{ar-1col}

\usepackage[numbers,sort&compress]{natbib}
\usepackage{amsmath}

\jname{Xxxx. Xxx. Xxx. Xxx.}
\jvol{AA}
\jyear{YYYY}
\doi{10.1146/((please add article doi))}

\newcommand{\myeq}[2][]{Equation#1~(\ref{#2})}
\newcommand{\myfig}[2][]{Figure#1~\ref{#2}}

\begin{document}

\markboth{Nicolas Bruot and Pietro Cicuta}{Colloidal Models of
  Hydrodynamic Synchronization}

\title{Realizing the physics of motile cilia synchronization with driven colloids}

\author{Nicolas Bruot$^1$ and Pietro Cicuta$^2$ \affil{$^1$Institut
    Lumi\`ere Mati\`ere, UMR5306 Universit\'e Claude Bernard Lyon
    1~--- CNRS, Universit\'e de Lyon, Institut Universitaire de
    France, 69622 Villeurbanne cedex, France} \affil{$^2$Cavendish
    Laboratory, University of Cambridge, Cambridge, UK, CB3 0HE}}

\firstpagenote{First page note to print below DOI/copyright line.}

\begin{abstract}
  Cilia and flagella in biological systems often show large scale
  cooperative behaviors such as the synchronization of their beats in
  ``metachronal waves''.  These are beautiful examples of emergent
  dynamics in biology, and are essential for life, allowing diverse
  processes from the motility of eukaryotic microorganisms, to
  nutrient transport and clearance of pathogens from mammalian
  airways.  How these collective states arise is not fully understood,
  but it is clear that individual cilia  interact mechanically,
  and that a strong and long ranged component of the coupling is
  mediated by the viscous fluid.
  We review here the work by ourselves and others aimed at
  understanding the behavior of hydrodynamically coupled systems, and
  particularly a set of results that have been obtained both
  experimentally and theoretically by studying actively driven
  colloidal systems. In these controlled scenarios, it is possible to
  selectively test aspects of the living motile cilia, such as the
  geometrical arrangement, the effects of the driving profile and the
  distance to no-slip boundaries. We outline and give examples of how
  it is possible to link model systems to observations on living
  systems, which can be made on microorganisms, on cell cultures or on
  tissue sections.  This area of research has clear clinical application in
  the long term, as severe pathologies are associated with compromised
  cilia function in humans.
\end{abstract}

\begin{keywords}
  hydrodynamic synchronization; driven colloidal particles; motile cilia;
  metachronal wave.
\end{keywords}
\maketitle

\tableofcontents

\section{INTRODUCTION}\label{sec:INTRO}

Biological systems are often found to make use of ``simple'' and
generic force-based mechanisms.  Since most of life exists in a fluid
state, an obvious medium for the transmission of force is the
hydrodynamic interaction: the flow fields transmit forces between
objects moving relative to each other or relative to the background fluid. At the microscopic cellular
level, velocities and length scales combine to give a very low
Reynolds number. Hydrodynamic coupling is an important feature in
various biological flows, with key consequences in apparently diverse
phenomena such as the motility of
microorganisms~\cite{Taylor1951,lauga09,lauga11,golestanian11b,gaffney11,goldstein15b,gompper15},
circulation in the brain~\cite{breunig10} and functioning of the
ear~\cite{kozlov11}.  In many cases motile cilia are involved in the
generation of the flows, and the fluid acts as a medium that will
couple the different beating cilia.  A whole ciliated tissue can
therefore be seen physically as a system of coupled oscillators.
Because of the complexity of the beat pattern of a motile cilium and
of its interaction with the surrounding fluid, simplified models of
coupled oscillators have been developed by us and by other groups to
address the question of the emergence of cooperative
behaviors~\cite{golestanian11b,gompper15}.  Such models have been
very useful to describe the features of the coordinated patterns as a
function of a limited number of control parameters.

{We discuss here
different models of oscillators that are used to gain insight in this
topic, through experiment and theory, particularly  to  develop our
   understanding of the link between micro- and macroscopic dynamics.
   For example there is evidence that clusters of close neighbours phase-lock into
   clear ``dynamical motifs''~\cite{cicuta12z}, and classifying the
   main motifs under biologically relevant conditions will show which
   parameters are (or may be) optimized to sustain synchronization and
   flow over large scales in the biological context.  Much remains
   to be explored in these well controlled models:  fluids like mucus are complex and
   viscoelastic~\cite{gheber94};
    flow occurs near soft tissues and with cilia close to each other,
   which is known to bring new hydrodynamic effects
     ranging from a shorter range coupling~\cite{stark11} to complex
     near-surface circulation patterns~\cite{vilfan10}.
      Experiments on the colloidal models have provided unexpected results
     for example on  the nontrivial consequence of thermal noise in coupled
     active systems~\cite{cicuta10a} and on the role of feedback frequency~\cite{bruot11}. }

The review is organized as follows.  Section~\ref{sec:CILIA}
introduces biological cilia, focusing on the origin of their activity
that ultimately leads to the generation of a net flow.  In
section~\ref{sec:colloids_physics} we present the theoretical
framework for the study of colloidal model oscillators, and in
section~\ref{sec:colloidal_oscillators} we describe the coarse-grained
models of oscillators that have been implemented experimentally with
particles driven by optical tweezers.
%
To compare predictions of the simplified colloidal models to natural
systems, it is relevant to also investigate real ciliated tissues and
model microorganisms.  This is discussed in
section~\ref{sec:sync_in_real_systems} with examples of increasing
complexity: synchronization of the two flagella of \emph{Chlamydomonas
  reinhardtii}, metachronal waves in \emph{Paramecium} and
\emph{Volvox}, alignment of cilia in mammalian mucociliary tissues and
observations on diseases affecting the respiratory tract.  Finally, in
section~\ref{sec:outlook} we present our view on the implications of
this work in the areas of clinical diagnosis of cilia-related
diseases, in opening new possibilities in microfluidics for flow
generation and control at the micrometric scale.
The overarching (and largely outstanding) question addressed in this
review is what determines the character of the dynamical steady state,
and the related corollary of what has gone ``wrong'' in clinical
conditions in which cilia coordination is observed to be compromised.

\begin{figure}[t!]
  \includegraphics{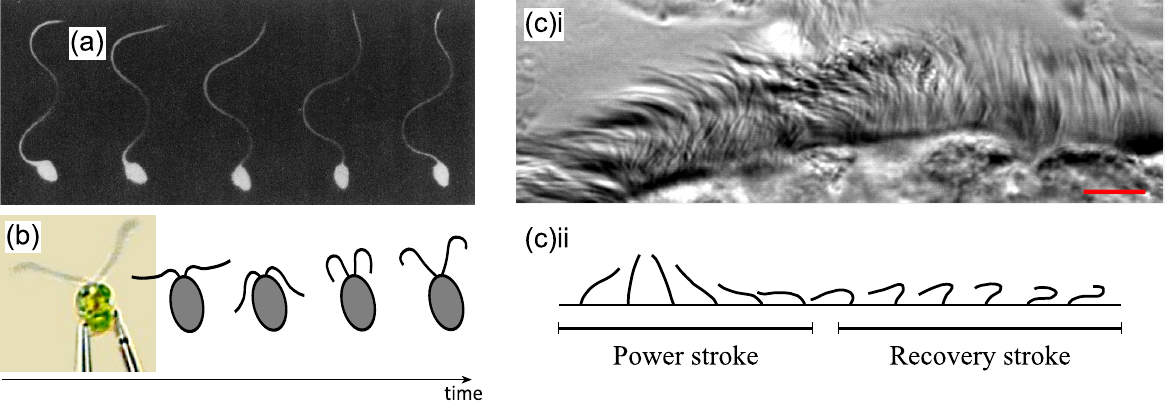}
  \caption{\label{fig:cilia_overview} Flagella and cilia.
    (a)~Flagellum of a sea urchin sperm cell showing a planar
    wave-like pattern~\cite{rikmenspoel85}.  The snapshots are taken
    at 10\,ms time intervals.  (b)~Biflagellated \emph{Chlamydomonas}
    cell, held on a micropipette, with clearly visible flagella
    (adapted from \cite{goldstein09b}). Sketches illustrate the
    ``breaststroke'' phase-locked motion which takes place for the
    majority of time and corresponds to directed motion. (c)i~Snapshot
    of side-on human airway epithelial cilia displaying a metachronal
    wave (scale bar: 5\,$\mu$m).  (c)ii~Sketch respresentation of the
    time evolution of the shape of a cilium as in (c)i.  The periodic
    cycle is made of a ``power stroke'', during which the cilium
    pushes the fluid, and a ``recovery stroke'', with the cilium
    coming back to its initial position at reduced drag.  In a
    multicilated system, in presence of a metachronal wave, this
    sketch also represents the spatial distribution of cilia shapes as
    a snapshot in time.}
\end{figure}

\section{CILIA}\label{sec:CILIA}

\subsection{Cilia structure and mobility}

\begin{figure}[t!]
  \includegraphics{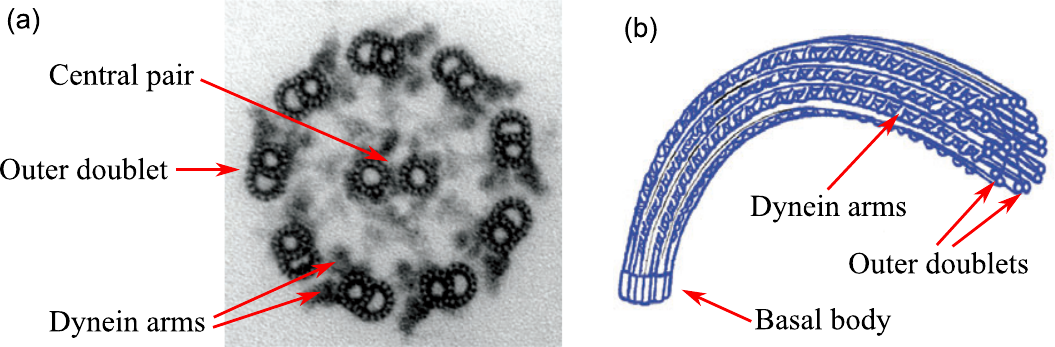}
  \caption{\label{fig:cilium_structure} The ``9+2'' internal structure of
    a motile cilium is widely conserved in eukaryotes.  (a) Axoneme cross-section showing nine
    microtubule doublets surrounding a central
    pair~\cite{lindemann10}.  (b)~Flagellar axoneme extending from the
    basal body on which the cilium is anchored to the rest of the
    cell.  The dynein arms are the motor of the cilium beat and induce
    the sliding of microtubules against each other (from neighbouring
    doublets), causing the overall bending of the
    filament~\cite{lindemann10}.}
\end{figure}

Motile cilia are remarkably complex organelles, evolved early in life,
whose structures is highly conserved throughout biology.  Only some
cells have motile cilia, either with the purpose of swimming, or to
generate a flow.  Mammalian sperm cells have a single, approximately
70\,$\mu$m long flagellum attached to the body
(\myfig{fig:cilia_overview}(a)), while the \emph{Chlamydomonas} alga
has two flagella, about 12\,$\mu$m in length~\cite{bayly10}
(\myfig{fig:cilia_overview}(b)).  In airway tissues
(\myfig{fig:cilia_overview}(c)), cells are multiciliated and cilia
form dense arrays (2400 cilia/cm$^2$~\cite{mercer94}); the filaments
are much shorter, on average 7\,$\mu$m for human
airways~\cite{leopold09}.  While the length of cilia and flagella can
vary significantly depending on the system, the internal structure of
both cilia and flagella in eukaryotic cells is very
conserved~\cite{vincensini11}, leading to a fairly constant diameter
of about 200\,nm.  In most of the cases, when looking at a
cross-section of a cilium, the internal structure (axoneme) consists
in 9 microtubules doublets arranged on a circle in primary cilia, and
9 microtubules doublets in circle that surrounds two central single
tubules in motile cilia~\cite{vincensini11}
(\myfig{fig:cilium_structure}).  The two structures are often called
``9 + 0'' and ``9 + 2'' respectively.  The microtubule doublets are
responsible for the movement of motile cilia, and the orientation of
the central tubules sets the plane of beating. In a cilium or
flagellum, each doublet is a pair of dynein arms that slide against
each other~\cite{bray00}. This causes the whole cilium to bend, as
shown in \myfig{fig:cilium_structure}(b). The energy required for the
sliding is provided by adenosine triphosphate (ATP). The details of
the motion of the filament are still being
debated~\cite{lindemann10,riedel-kruse07}, but it is well known that
the outcome of all the activity, from the molecular up to the single
cilium level, is an oscillating filament.  Typical cycles are shown
in~\myfig[s]{fig:cilia_overview}(a), (b) and (c)ii.  For the short
cilia in (b,c), the cycle starts with the ``power stroke'' with a
fairly straight filament that changes its orientation, thus pushing
the fluid. It returns to its initial position curled up, in the
``recovery stroke'' thus minimizing the interaction with the fluid. In
the two strokes, a cilium has a very different drag coefficient
because of the different conformations; this asymmetry makes the cycle
non reciprocal, a necessary condition to provide a net force on the
fluid over a period in a low Reynolds number environment. Cilia
typically beat at a frequency of 5 to 100\,Hz~\cite{satir07}. Other
beating cycles exist.  For example, in uniflagellar spermatozoa, the
cycle can be a wave-like planar (as in \myfig{fig:cilia_overview}(a))
or a helical pattern that propagates along the
filament~\cite{bray00,woolley01,gaffney11}, pushing the cell
forward. The planar pattern in \myfig{fig:cilia_overview}(a) is
somehow similar to the S-shape slithering motion of snakes, although
the problem is more complex since the fluid is not
immobile~\cite{Taylor1951}. Attempts have been proposed to account for
the mechanisms at work within a cilium~\cite{hilfinger08,hilfinger09},
and to reproduce the cycle of oscillation~\cite{brokaw09,brokaw09a}.
Already at the level of a single cilium there are complex non-linear
interactions amongst the constitutive components.

Physical considerations on cilia can be made on a single cilium.  How
the cilia units are structured and function is itself a fascinating
question, which is well understood at the molecular scale, and is
being addressed at the ``system'' level in terms of filament
dynamics~\cite{hilfinger08,lauga09,gaffney13}.  However, a full model
that includes in general the internal motor forces and cilia transfer
of momentum to the surrounding fluid does not yet exist.

\subsection{Cilia coordination: Biological and physical views}

The function of motile cilia is essential in life: In complex
organisms such as mammals, cilia are expressed in tissues to drive
circulation in the brain, airways and fallopian tube. In these organs
the active cilia units interact with each other creating a coupled
dynamical system, which can lead to coordinated beating in a
``metachronal wave''~\cite{gueron97,gompper13}, see
\myfig{fig:cilia_overview}(c)i.  This wave is particularly
efficient at sustaining directed surface flow~\cite{bray00}. Nearby
cilia will phase-lock, and may for example beat in-phase or out of
phase, or indeed may be in a condition where they can readily switch
between the two dynamical states as in the algae \emph{Chlamydomonas
  reinhardtii}~\cite{goldstein09a}.

Coordinated motion is crucial for the effective functioning of cilia
and flagella, the elements of eukaryotic cells implicated in
generating fluid flows and
motility~\cite{bray00,gompper13,wallingford14}, and may be exploited
in artificial conditions~\cite{tlusty06,hanggi09b,vilfan10,dogic11}
and low Reynolds number (Re)
swimmers~\cite{yeomans09,cicuta09a,palagi13}.  Motile flagella and
cilia interact through the velocity field in a low Re
regime~\cite{gueron97,lenz08,lauga09}.    The magnitude of the coupling forces
is comparable to the random thermal forces, raising the question of
how biological systems might exploit this competition between locking
and random phase drift, or how architectures leading to stable
synchronized states of multiple cilia can have
evolved. Synchronization is a general phenomenon in
nature~\cite{pikovsky01}, with many technological applications.
Hydrodynamic synchronization at low Re is a well defined subset of this, in
which coupling is viscous~\cite{lauga09}, noise (thermal and intrinsic
contributions) is important~\cite{goldstein09a,ma14}, and the details
of the driving forces play a key role~\cite{lenz08,stark11}. Recent
progress in hydrodynamic synchronization is reviewed
in~\cite{golestanian11b,gompper15}, and an overview of low Reynolds
number (Re) flows is in~\cite{lauga09}.

\begin{figure}[t!]
  \includegraphics{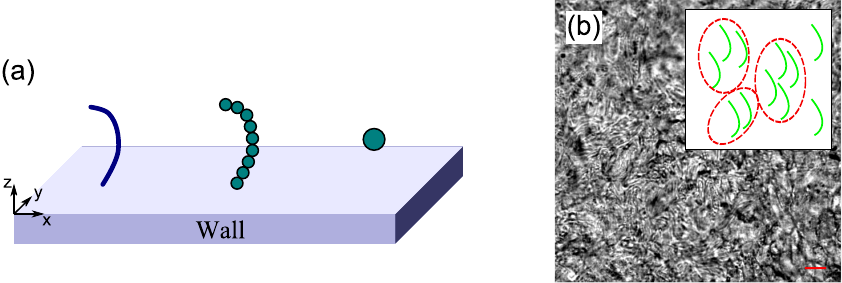}
  \caption{\label{fig:models_and_expectations}  Hydrodynamic coupling
leads to coordinated motion of
   cilia and can be studied using simpler models.  (a)~In a
    coarse-grained model, a single cilium can be modeled as a chain of
    spherical colloidal particles, and even as a single bead.  Despite
    their simplicity, these descriptions approximate well (in the far
    field limit) the characteristics of the fluid flow generated by a
    full filament.  (b)~Top view of a culture of ciliated cells (scale
    bar: 10\,$\mu$m).  Movies of such cultures can be studied to look
    for local dynamical patterns (illustrated in the inset) that are
    predicted by model studies involving the hydrodynamic coupling.}
\end{figure}

One approach to understand hydrodynamic synchronization is to study
systems of a few flagella.  For example, the biflagellated alga
\emph{Chlamydomonas} (\myfig{fig:cilia_overview}(b)) has been a
test-bed for experiments exploring synchronization of
flagella~\cite{goldstein15b}, recently showing that hydrodynamic flows
are sufficient to describe phase-locked cilia
dynamics~\cite{goldstein14}.  Particle-based experimental and
conceptual models, where the emerging dynamics in coupled arrays of
driven colloidal phase oscillators can be probed, have shown clearly
the role that geometry~\cite{cicuta12,cicuta12z,bruot12b,bruot13a},
type of drive~\cite{golestanian11a,cicuta12t,dileonardo12}, and
details of the driving
potential~\cite{stark11,golestanian12,cicuta12t,bruot13b,dileonardo13}
can have on the collective state.  By enabling selective control of
individual parameters, and analytical results in some limit
conditions~\cite{golestanian11a,golestanian12,cicuta12}, these models
form a solid foundation to build our understanding of natural systems
such as mucociliary tissues.  These models maintain the correct
far-field form of the hydrodynamic flow caused by a cilium, can
describe the presence of a nearby solid surface, and most importantly
can account for the physiological properties of the beating cilia
(i.e. shape of the stroke, with power and recovery phases) by matching
the properties of living system cilia cycles as the driving rule.

\begin{textbox}
  Most eukaryotic cells have cilia, which may be motile or
  immotile. It is now thought that most cilia can act as chemical- or
  mechano-sensors, while only motile cilia enable epithelial cells in
  various tissues to generate surface flows~\cite{fliegauf07}. Motile
  cilia are highly conserved structures in the evolution of organisms,
  essentially the same in plants and animals; they generate the
  transport of fluid by periodic beating, through remarkably organized
  behavior in space and time. It is not clearly known how these
  spatio-temporal patterns emerge and what sets their properties.
  Individual cilia are nonequilibrium systems with many degrees of
  freedom, but can be represented by simpler effective force laws that
  drive oscillations, and paralleled with nonlinear phase oscillators
  often studied in physics.  Recent models capture the key physical
  elements at work within each filament~\cite{hilfinger09}: each
  cilium undergoes periodically a forward ``power stroke'', followed
  by a backward ``recovery stroke'' in which it is more bent and
  closer to the surface, thus resulting in a net momentum transfer to
  the fluid over a cycle (\myfig{fig:cilia_overview}(c)ii).  The
  source of synchronization observed in cilia and flagella is often
  thought to be of hydrodynamic origin, based on many
  observations~\cite{gueron97,tuval08,goldstein09b,goldstein14}, but
  the details of coupling in the case of mucociliary tissues (and
  specifically of the airways) are not known and are a topic of
  current research.  Synchronization in systems of many oscillators
  holds many open questions, and is close to frontiers of soft matter
  and statistical physics dealing more widely with the collective
  behavior of active elements, and the emergence of non-equilibrium
  structures~\cite{prost13,ramaswamy13}.
\end{textbox}

A fruitful approach to date has been to build experimental models of
driven phase oscillators, sufficiently simple for both theoretical and
experimental studies.  Experimentally, these involve optically trapped
colloidal sized spheres, which play the role of beating cilia,
maintaining the same regime of length and time scales
(\myfig{fig:models_and_expectations}(a)).  In an attempt to model
(both experimentally and theoretically) the physics of hydrodynamic
synchronization of cilia, two main ideas have emerged.

One model consists of a ``geometric switch'': the model cilium has two
states corresponding to its back and forth motions.  It displays a
discontinuity in the driving force and is motivated by the existence
of the power and recovery strokes (\myfig{fig:cilia_overview}(c)ii) and
by the fact that molecular motors undergo discrete
attachment/detachment events which couple to the force generation.  It
was first studied theoretically
in~\cite{gueron99,bassetti02,bassetti03}, then by our
group~\cite{cicuta10a,cicuta12,cicuta12z}, and others~\cite{stark11}.

In contrast with the configuration-controlled system above, a
stress-controlled ``rotor'' model has also emerged~\cite{lenz08}.
This has recently been studied very generally for two rotors by Uchida
and Golestanian~\cite{golestanian11a,golestanian12}, extending the
case of circular orbits by Lenz~\cite{lenz08}.  A series of important
results has been obtained modelling cilia as
rotors~\cite{vilfan06,lenz08,golestanian11a}; in recent
work~\cite{golestanian12,bruot13b} it was shown that both the
flexibility of the orbit, and the mean force profile, determine the
strength of synchronization.  By small adjustments of its physical
parameters, it is possible to achieve control over the state of
synchronization of the system.

Some of our results on the two models are discussed in
Section~\ref{sec:colloidal_oscillators}.  It is not yet established
which of the two ideas (stress- or configuration-controlled
oscillators) is most appropriate to describe a biological scenario.

The research frontier is now to develop the model experiments further,
to include key biological aspects (such as the high density of cilia
in some organs, or the presence of a non-Newtonian fluid), and to
enable rules and properties at the micro-scale to be linked with the
macroscopic dynamics.  In this sense, understanding cilia
synchronization in biological systems is an open and very important
area of research. The recent progress gained in simple systems,
coupled with advances in optical microscopy, ciliated cell culture,
microfluidics, optical manipulation and camera technology, make this
area ripe for experimental investigation.


\subsection{What can we learn from models: Relating the microscopics
  to collective dynamics}

Each biological cilium is itself a complex structure (see \myfig{fig:cilium_structure}), and its own
regular beating (and switching) is constrained by a mechanical
feedback~\cite{riedel-kruse07,hilfinger09}.  In a given flow
condition, the mechanical stress and the geometrical configuration are
coupled parameters, and the feedback condition of the geometric switch
is a simple way to account for how a cilium senses the moment to
switch between so-called power and recovery
strokes~\cite{gueron97}. It is clear that a complex scenario such as
this requires models capable of joining up quite different length
scales, from molecular to cilium level, and up to the interactions of
cilia with the fluid and each other.  In this spirit, and depending on
the focus of the question, some aspects have to be suitably simplified
and reduced.

Certain organisms (biological model systems) also allow, to some degree,
to address specific aspects of such an entangled problem.  For the
alga \emph{Chlamydomonas}, the synchronization of the beating
filaments was shown
recently~\cite{goldstein09b,goldstein11,friedrich12} and the
visualization of cilia shape was also achieved~\cite{bayly10}.  While
\emph{Chlamydomonas} has been very useful for developing and tuning
physical models to biological systems, ultimately important goals lie
in understanding human physiology, where ciliated tissues have large
numbers of cilia, and cilia arrangement is likely to be important. In
tissues such as the trachea of mammals, the average frequency of
beating and speed of the wave can be measured readily, but observation
of the individual beating cilia is hampered by the presence of
optically opaque mucus~\cite{wong93,chilvers00}, see
\myfig[s]{fig:cilia_overview}(c)i and \ref{fig:cilia_side_view}(b).

In this context, colloidal model experiments, where the parameters of
the system can be tuned readily and visualization is straightforward,
have an important role to play in elucidating the factors controlling
emergent collective states. Colloidal particle driven systems (in a
way, a model even simpler and more flexible than \emph{Chlamydomonas})
can be used to demonstrate specific effects: for example thermal noise
(capable of course of destroying the synchronization if sufficiently
large) introduces a delay between the coupled
oscillators~\cite{cicuta10a}.  The conditions for phase locking can be
predicted in simple scenarios~\cite{cicuta12t}, and we have developed
design principles to explain the emergence of characteristic dynamical
patterns in small networks of oscillators~\cite{cicuta12,cicuta12z}.
The systems we investigated to date can be thought of as idealized
cilia synchronization experiments: The optical trap force plays the
role of the molecular motors that induce active movement, and the
hydrodynamic flow field produced by the moving beads well represents
beating cilia, at least at large distances~\cite{lauga09}.
These models are also general experimental systems to probe the
physics of stochastic and actively forced hydrodynamically coupled
oscillators.
From the theoretical perspective, working with models enables to
develop well defined general results, for example various models have
been proposed recently addressing different aspects of metachronal
waves~\cite{lenz08,hilfinger09,guirao07,lauga09,stark11,friedrich12}.

The role of models, where individual parameters can be tuned
selectively, is to establish the general rules and links between the
emergence and the properties of synchronized states (e.g. the
collective phenotypes, like wave dynamics and the spatial or temporal coherence of coordination) and the key physical properties of the
 individual active structures (e.g. mechanics, activity and
geometry of each motile cilium, and fluid rheology).

\section{PHYSICS OF DRIVEN COLLOIDS}
\label{sec:colloids_physics}

To gain an understanding in the synchronization of cilia and the
emergence of metachronal waves, it is necessary to coarse-grain the
filaments in various ways, and analyze the emergent behavior
experimentally, numerically or with analytic calculations.  In a quite
accurate description of the drag with the fluid, the filament can be
replaced by a chain of spherical beads or rods
\cite{vilfan10,coq11,gompper13,khaderi12}
(\myfig{fig:models_and_expectations}(a)). This can also be realized
experimentally with magnetic particles self-assembling in
chains~\cite{vilfan10,vilfan12,dreyfus05}.  These models are useful to
study the flow generated close to the filament and to obtain its shape
during a beating cycle, but so far it has only been possible to
actuate the filaments with a uniform external force.  Elliptical
particles have also been used~\cite{kavre15}.  To consider questions
relating to synchronization between different cilia, an even more
coarse-grained model can be used: Since the filament can be seen as a
point in the far field, it can be assumed that the fluid flow
generated by the cilium is the same as the one created by a moving
sphere. Hence, a cilium can be modelled as just a sphere,
see~\myfig{fig:models_and_expectations}(a).  Many recent models use
this level of coarse-graining~\cite{lenz08,vilfan06,golestanian11a},
as this greatly simplifies the calculation of drag forces, both those
acting on the individual object, and the force induced by one object
on another. The cycle shape and bead velocities contain in a
coarse-grained fashion the cilium properties, and are at the core of
the model. It is possible, as outlined in section~\ref{sec:colloidal_oscillators}, to deploy these models
(and also carry out experiments) without locking the phase of each
cilium to an external driving clock.

\subsection{Hydrodynamic interaction}\label{sec:sub:hydrodynamics}
This section recalls the main equations that govern colloidal models,
and the numerical methods that can be used to simulate systems of
hydrodynamically coupled oscillators.

One key interest of colloidal oscillators as a model of cilia is to
provide simple ways to describe the hydrodynamic interaction.  Studies
until now we have assumed a Newtonian fluid in a low Reynolds' number
configuration, that the propagation of the transfer of momentum from
the fluid is instantaneous, and that the colloidal particles do not
rotate.  For a set of $N$ colloidal particles, the hydrodynamic
coupling between the external forces acting on the colloids (all
forces but the one from the fluid) and the resulting velocities of the
particles is given by a relation of the form
\begin{equation}
  \label{eq:mobility_matrix}
  \mathbf{v}_i = \sum_{j = 1}^n \boldsymbol{\mu}_{i,j} \mathbf{F}_j \ .
\end{equation}
where $\mathbf{v}_i$ is the velocity of bead $i$, $\mathbf{F}_j$ the
external force acting on bead $j$, and $\boldsymbol \mu$ is an $N
\times N$ mobility matrix whose elements are $3 \times 3$ second-rank
tensors.  The choice of $\boldsymbol \mu$ depends on the presence or
not of no-slip boundary conditions (a nearby wall) and on the level of
approximations regarding the finite radius $a$ of the particles (near
field effects).  We outline below the most commonly used mobility
matrices.

\subsubsection{Far field, no wall: Oseen tensor}

A moving particle in an unbounded quiescent fluid generates a flow
$\mathbf{u}(\mathbf{r})$ as a result of the external forces acting on
the particle.  Such a solution of the Stokes
equation~\cite{dhont96a,brenner83} for the fluid is called a
\emph{Stokeslet}~\cite{chwang75}.  If a second particle is placed at
$\mathbf r_i$ in the fluid it will simply follow the fluid at the
velocity $\mathbf u (\mathbf r_i)$.  In the case of $N$ particles,
when the interparticle distances $r_{i,j} = |\mathbf r_i - \mathbf
r_j|$ are large compared to their radius $a$, and when keeping only
the leading orders in $a/r_{i,j}$, the mobility matrix simplifies to
the Oseen tensor $\boldsymbol \mu = \mathbf H$.  In that case, the
diagonal terms simply represent the Stokes drag on a sphere, and the
non-diagonal terms are directly related to the
Stokeslets~\cite{edwards86,dhont96a}:
\begin{equation}
  \mathbf{H}_{i,j} = \left\{
  \begin{array}{lll}
    \dfrac{1}{8 \pi \eta r_{i,j}}
    \left( \mathbf{I} + \mathbf{\hat{e}}_{i,j}
      \mathbf{\hat{e}}_{i,j} \right) &
    \text{if} & i \neq j\\[5mm]
    \dfrac{\mathbf{I}}{6 \pi \eta a} & \text{if} & i = j \\
  \end{array}
  \right.
\end{equation}
with $\mathbf I$ the unit tensor and $\mathbf{\hat{e}}_{i,j}$ a unit
vector in the direction defined by the two particles $i$ and $j$.

\subsubsection{Near field, no wall: Rotne-Prager tensor}

A more accurate description of the hydrodynamic interaction than the
Oseen tensor accounts for the finite size of the particles in the
non-diagonal terms of the mobility matrix.  The correction, which
appears as a higher order term in $a/r_{i,j}$ takes into account that
a flow created by a particle is ``reflected'' on the other particles,
thus affecting the velocity of the first
particle~\cite{dhont96a,brenner83}.  To stay consistent with the
expansion to the next order in $a/r_{i,j}$, it is also necessary to
expand the Stokeslet flow of a single particle to the third order, as
well as taking into account that a particle moving in an non-uniform
flow undergoes a translational motion slightly different from the sole
fluid velocity at the position of the particle (\emph{Fax\'en theorem}
for translation).  All together, the corrections above lead to the
Rotne-Prager mobility matrix~\cite{dhont96a}
\begin{equation}
  \boldsymbol \mu_{i,j}^{\text{RP}} = \left\{
  \begin{array}{lll}
    \dfrac{1}{6 \pi \eta a} \left[ \dfrac{3 a}{4 r_{i,j}}
      \left( \mathbf I + \mathbf{\hat e}_{i,j} \mathbf{\hat
          e}_{i,j} \right) + \dfrac{1}{2} \left( \dfrac{a}{r_{i,j}} \right)^3
      \left( \mathbf I - 3 \mathbf{\hat e}_{i,j} \mathbf{\hat
          e}_{i,j} \right) \right] & \text{if} & i \neq j\\[5mm]
    \dfrac{\mathbf{I}}{6 \pi \eta a} & \text{if} & i = j\\
  \end{array}
  \right. \ .
\end{equation}

\subsubsection{Presence of wall: Blake tensor}

Many biological systems involve the hydrodynamic coupling between
objects that are moving close to a wall, with a no-slip boundary
condition.  This is particularly true for ciliated tissues, where the
filaments are anchored at the surface of the cells.  The Blake tensor
describes the interaction between spheres in a semi-infinite fluid
with a no-slip boundary condition at the surface.

For the diagonal terms of the mobility matrix, the presence of the
surface modifies the Stokes drag that can be expanded as a series in
$a / z_i$, where we take a system of coordinates $(x,y,z)$ with
$\mathbf{\hat e}_z$ normal to the surface and $z = 0$ at the
surface~\cite{perkins92}.  An expansion to the sixth order has already
been used to study the flow generated by arrays of artificial
cilia~\cite{vilfan10}.  The corresponding diagonal terms read:
\begin{equation}
  \left\{
  \begin{array}{ll}
    \boldsymbol \mu_{i, i}^{x, x} &= \boldsymbol \mu_{i, i}^{y, y}
    = \dfrac{1}{6 \pi \eta a} \left[ 1 - \dfrac{9 a}{16 z_i} +
      \dfrac{1}{8} \left( \dfrac{a}{z_i} \right)^3 - \dfrac{1}{16}
      \left( \dfrac{a}{z_i} \right)^5 \right] \\[5mm]
    \boldsymbol \mu_{i, i}^{z, z} &= \dfrac{1}{6 \pi \eta a}
    \left[ 1 - \dfrac{9 a}{8 z_i} + \dfrac{1}{2}
      \left( \dfrac{a}{z_i} \right)^3 - \dfrac{1}{8} \left(
        \dfrac{a}{z_i} \right)^5 \right]
        \\[5mm]
        \boldsymbol \mu_{i, i}^{\alpha, \beta} &= 0 \text{ for }\alpha
        \neq \beta \ .
  \end{array}
  \right.
\end{equation}

For the non-diagonal terms, Blake proposed in~\cite{blake71} to
describe the fluid flow created by a Stokeslet near a surface by an
image method (as in electrostatics).  In this method, the no-slip
boundary condition at the wall is satisfied by describing the effect
of the wall as equivalent to an infinite fluid, but with a second
Stokeslet at the mirror position of the first Stokeslet and with an
opposite force.  For $N$ particles, this leads to the following
expressions for the Blake mobility matrix~\cite{blake71,pozrikidis96}:
\begin{equation}
  \boldsymbol \mu_{i,j}^{\text B} = \dfrac{1}{8 \pi \eta} \left[ \mathbf
    G^{\text S} (\mathbf r_i - \mathbf r_j) - \mathbf G^{\text S}
    (\mathbf r_i - \overline{\mathbf r}_j) + 2 z_j^2 \mathbf G^{\text
      D} (\mathbf r_i - \overline{\mathbf r}_j) - 2 z_j \mathbf
    G^{\text{SD}} (\mathbf r_i - \overline{\mathbf r}_j)\right] \ .
\end{equation}
with $\mathbf r_i = (x_i, y_i, z_i)$, $\overline{\mathbf r}_i = (x_i,
y_i, -z_i)$ and with the elements of the Green functions given
by~\cite{blake71,pozrikidis96}
\begin{equation}
  \begin{array}{ll}
    \mathbf G^{\text S}_{\alpha, \beta} (\mathbf r) &=
    \dfrac{\delta_{\alpha, \beta}}{r} + \dfrac{r_\alpha r_\beta}{r^3}
    \\[3mm]
    \mathbf G^{\text D}_{\alpha, \beta} (\mathbf r) &= (1 - 2
    \delta_{\beta, z}) \dfrac{\partial}{\partial r_\beta} \left(
      \dfrac{r_\alpha}{r^3} \right) \\[3mm]
    \mathbf G^{\text{SD}}_{\alpha, \beta} (\mathbf r) &= (1 - 2
    \delta_{\beta, z}) \dfrac{\partial}{\partial r_\beta} \mathbf
    G_{\alpha, z}^{\text S}(\mathbf r)
  \end{array}
\end{equation}
with $\delta$ the Kronecker delta.

In the most general case, when the size of the particles $a$ is not
negligible compared to $r_{i,j}$ or $z_i$, the Blake tensor can be
corrected for the beads radii.  In a derivation similar to the
Rotne-Prager tensor, the Fax\'en theorem
gives~\cite{gauger09,vilfan10}
\begin{equation}
  \boldsymbol \mu_{i,j} = \left( 1 + \dfrac{a^2}{6} \nabla^2_{\mathbf
      r_i} \right) \left( 1 + \dfrac{a^2}{6} \nabla^2_{\mathbf r_j}
  \right) \boldsymbol \mu_{i,j}^{\text B} \ ,
\end{equation}
for $i \neq j$.  The expanded version of this formula is heavy
and is not printed here but can be found in the supplementary
information of~\cite{vilfan10}.

\subsection{Thermal fluctuations}

As a consequence of the small scales involved, small biological
systems are subject to Brownian motion.  The random force $\mathbf f
(t)$ that acts on a spherical colloid is in practice well approximated
by white noise and is characterized by~\cite{quake99}:
\begin{equation}
  \left\{
    \begin{array}{l}
      \left< \mathbf f(t) \right> = 0\\
      \left< \mathbf f(t) \cdot \mathbf f(t') \right> = 2 \nu \: \gamma
      k_B T \: \delta(t-t') \ ,
    \end{array}
  \right.
\end{equation}
with $\gamma = 6 \pi \eta a$, and $\nu \in \{1, 2, 3\}$ the
dimension of the system.

In a system of several particles, this thermal fluctuating force has
to be included in the forces $\mathbf F_j$ of the mobility matrix in
\myeq{eq:mobility_matrix}.  It then leads to coupled terms of
noise in the velocities $\mathbf v_i$~\cite{quake99,cicuta10c}, which
we have shown to induce non-trivial correlations in motile
systems~\cite{cicuta10a}.  Thermal noise also induces rotation of the
particles and this rotation couples to the
translation~\cite{reichert04}.

\subsection{Numerical methods}

Numerical methods have provided a lot of insight into questions of
fluid dynamics, beat efficiency, and generally in finding solutions
for any model dynamical system that has been
posed~\cite{lenz08,gompper13,golestanian10a}, since these can usually
be treated analytically only in very special
cases~\cite{cicuta10c,golestanian11a,bruot11}.

 Brownian Dynamics
(BD) codes include the treatment of thermal fluctuations.  Such simulations  are extremely useful to provide quantitative
solutions to compare with experiments, and to extend theory approaches
(which are often simplified, and typically neglect thermal noise). BD
codes can be pushed to very large systems of over 100 fully coupled
model cilia (this is fewer than the active elements in multiciliated
tissues, but much larger than possible with any optical trap device),
even with a standard desktop computer.  Simulations of course assume
that the hydrodynamic interactions are known mathematically: The
equations outlined above do accurately describe coupled colloidal
particles in Newtonian liquids, but the problem becomes more
challenging when e.g. viscoelasticity is introduced.

In our own work we use a C++ BD program that implements the Ermak
McCammon algorithm~\cite{ermak1978}, providing a numerically fast way
to integrate the equations of motion of $N$ particles interacting with
the Oseen or Rotne-Prager tensors, with coupled thermal fluctuations. {In this
algorithm, momentum transfer is instantaneous (global) and noise has long-range spatial correlations.}

\section{COLLECTIVE DYNAMICS AND METACHRONAL WAVE MODEL EXPERIMENTS}
\label{sec:colloidal_oscillators}

Understanding the synchronization of cilia and flagella can be done by
means of \emph{in vitro} experimental observations, or by working on
simplified physical models that describe the beating pattern of
cilia. Since it is increasingly accepted that the hydrodynamic
interaction plays a key role in the synchronization of these
systems~\cite{gueron97,gueron99,gheber94,guirao07}, the models should
include that coupling.  This section focuses on colloidal models where
the hydrodynamic interaction dominates.  Other interactions, such as
mechanical~\cite{geyer13} through the cells, or steric repulsion might
also be relevant.

The geometric switch and rotor models below are simple enough to carry
out numerical simulations or analytical calculations by reducing the
multiple degrees of freedom of the complex systems into a few control
parameters which can be tuned.  These help to understand theoretically
the biological systems.  They also make it easier to account for the
hydrodynamic interaction.  For example, sperm flagella have been first
modelled by a waving sheet~\cite{Taylor1951,elfring09} or by a 2d or
3d filament~\cite{gray55,lighthill76,gueron92}, in order to determine
the fluid flow they create.  Experimental realizations of actuated
filaments that can mimic the motion of a cilium also exist, at the
micrometric~\cite{vilfan10,coq11,keissner12}, and
macroscopic~\cite{sareh12} scales.

\subsection{Experimental realizations of colloidal phase oscillators}

The colloidal particle experiments are based on the physical intuition
that the complex structural details of the cilia might be
coarse-grained into the details of how the colloidal particles are
driven.  Two classes of colloidal models (geometric switch and stress
driven) have been developed in this vein, exploring in first instance
the conditions for optimal coupling.  We have also introduced a
modified version of the geometric switch model with a free orientation
so that collective alignment properties could be investigated.  We
present here these coarse-grained models and their most interesting
collective features that have been observed by us and other groups.

\begin{figure}[t!]
  \includegraphics{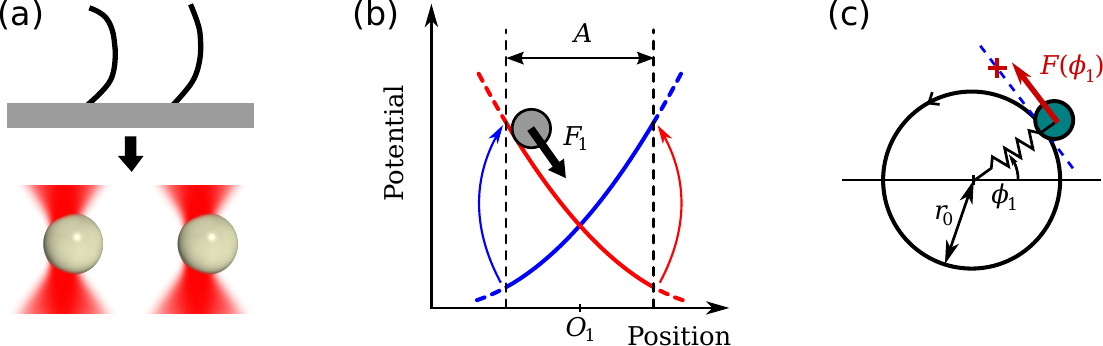}
  \caption{\label{fig:colloidal_models} Colloidal model oscillators.
    (a)~Our experimental approach to the cilia synchronization
    problem: Each cilium is represented by a spherical
    micron-sized particle,  driven by optical tweezers in a way that
    represents the cilium beat and contains cilia properties in a coarse-grained fashion.
    Two classes of models have been studied extensively in theory and experiment, running multiple coupled phase-free
    oscillators  to study their hydrodynamic synchronization.
    (b)~Geometric switch ``rower'' oscillator: Two traps are defined (red and
    blue potentials) and are switched on and off in turn.  When the
    trap on the right is on, the bead moves to the right, and the
    traps are switched when the particle reaches the position
    indicated by the right dashed line.  It is then pushed to the left,
    and a second geometric boundary (left dashed line) defines the
    next trap switch.  (c) ``Rotor'' model: A path is chosen for the
    particle (circular here).  The colloid is driven by a trap that is
    maintained at a controlled distance $\epsilon(\phi_1)$ ahead of
    the particle, and on the tangent to the circle at the bead.  The
    tangent force $F(\phi_1)$ is therefore controlled; the phase
    $\phi_1$ is determined from the bead position.  In addition, the
    bead may deviate from the circular path in the radial direction,
    and a harmonic restoring force of controlled stiffness
    (represented by a spring on the sketch) tends to keep the particle
    on the predefined path.}
\end{figure}

Experimentally, optical tweezers can be used to impose external forces
(in the range from fractions to hundreds of piconewtons) onto
colloidal particles (dielectric spheres of size range between half and
5 micrometers, \myfig{fig:colloidal_models}(a)). In purpose-made
optical tweezers setups like~\cite{bruot13b} it is possible to generate
many laser beams (or achieve many traps by time sharing one beam
across multiple positions, at much higher frequency than any dynamics
in the system) by deflecting beams via surface light modulators or
acousto-optical modulators. By analysing the positions of the
colloidal particles at high frame rate (typically hundreds of times
per second), and implementing a feedback loop, it is then possible to
make phase oscillators, i.e. to drive colloids cyclically in such a
way that the phase of each one is not set externally.  In
geometric switch models, the update of the laser trap positions is
decided based on the instantaneous configuration of the particles; in
stress-controlled orbits, the laser trap ``driving'' each colloid is
maintained a predefined distance ahead of the colloidal particle, on a
predefined track.


Within each class of model (configuration- or stress-controlled), many
biologically and medically important elements can be tested: the role
of fluid rheology, the vicinity to solid surface boundaries, the
robustness to external flows and perturbations in the flow, the
effects of heterogeneity in cilia spatial distribution, and the
robustness to loss of cilia in patches.

\subsection{Geometric switch: The ``rower'' model}
\label{sec:rowers}

\subsubsection{The model}

In the experiments of~\cite{cicuta10a}, two beads are confined in
separate harmonic wells (\myfig{fig:colloidal_models}(a)).  The
position of each well is linked to the spatial configuration of the
beads via a ``geometric switch''. Specifically, the laser trap on each
particle is moved between two positions a distance apart, as shown in
\myfig{fig:colloidal_models}(b), following the rule that the switch of
trap position is triggered when a particle approaches to within a
distance from the minimum of the active potential. This
feedback-controlled motion of the traps is sufficient to induce
sustained oscillations, and each particle undergoes long-time periodic
motion with a fixed amplitude. Crucially, when more than one bead is
present in the system, the geometric switch is determined
independently for each bead (200 times per second in the current
setup), so that the external trap forces do not themselves impose the
phase of oscillation nor its period.  Since the bead radius is
typically a few micrometers and the trap stiffness is between 1 and
100\,pN/$\mu$m, with the typical viscosity of 1\,mPa$\cdot$s or above,
in the absence of other external forces, a particle in a harmonic
potential undergoes overdamped stochastic motion driven by thermal
forces~\cite{quake99}.
%
%
We more recently explored variations of this model such as two-state
oscillators driven by non-harmonic potentials~\cite{bruot11,cicuta12t}.


\subsubsection{Synchronization of rowers}

In our first experiments~\cite{cicuta10a}, we showed that two driven
oscillating colloidal spheres in harmonic potentials lock into
antiphase motion, exhibiting a surprising behavior caused by the
interplay of thermal noise and hydrodynamic interactions, as well as
general features typical of coupled nonlinear oscillators such as Arnold tongues.

The shape of the driving potential also affects the state of
synchronization.  Wollin and Stark showed this in the geometry of a
linear chain of two or more oscillators, studied numerically in the
absence of noise~\cite{stark11}.  For two driven particles, we found
experimentally that the curvature of the potential does indeed
determine the stable synchronized state: in-phase or in
antiphase~\cite{cicuta12t}.  More specifically, harmonic-like
potentials with a force that decreases when approaching the center of
the well lead to antiphase synchronization, while an increasing force
generates in phase synchronization.  This result is illustrated by
\myfig{fig:main_results}(a).  Here, the parameter $c$ characterizes
the ``curvature of the potential'' --- with constant driving forces
corresponding to $c = 0$ --- and $\left< Q \right>$ is an order
parameter that takes a value of $-1$ for in-phase synchronization and
$+1$ for antiphase motion.  These experiments show a transition of
the synchronized state, and were backed up by BD simulations with the
coupling described by the Oseen tensor.  At $c \approx 0$ (non-curved,
linear potentials), the two rowers do not synchronize even at low
temperature, which is a manifestation of the fact that breaking
time-reversal symmetry is a necessary condition for the system to
synchronize.  Analytical considerations are also possible at this
level of description: a theory based on the solving of the equations
of motion in the presence of thermal fluctuations (the thermal
noise leads to distributions of the particle first-passage times that
trigger a trap move) successfully approximates the mean order parameter.

The role of noise in these systems is a relevant question given the
micrometric size of the system (Brownian motion) and presence also of  bio/chemical noise
due to the molecular processes in the living systems.  We have highlighted that thermal
noise can prevent synchronization~\cite{bruot11,cicuta12t} and trigger
phase slips when the two oscillators have different intrinsic
frequencies~\cite{cicuta10a,bruot11}.

The position of the ``model cilia'' is also very important: we looked
experimentally at arrays of $N=3$, 4 and 5 oscillators (and more
general systems, numerically)~\cite{cicuta12,cicuta12z}, making the
surprising discovery that while polygonal arrays of 4 or more colloids
behave like the 2-particle system, and synchronize with the nearest
neighbors in antiphase (for harmonic drive), a system of 3 equally
spaced colloids synchronizes in-phase. Other odd-number systems share
the property of having phase-locked travelling waves clockwise and
anticlockwise.  Studying these small networks, we realized that the
non-equilibrium dynamical steady-state is predominantly formed by the
eigenmode with longest relaxation time when the driving potentials
have positive curvature (e.g. harmonic wells with a positive
coefficient).  On the contrary, the eigenmodes with the shortest
relaxation times dominate the dynamics in systems where the curvature
is negative, because modes are growing and the shortest time constant
modes grow the fastest.  Therefore, for a given form of the coupling
tensor, the character of the collective dynamics can be predicted from
the spatial configuration, and the knowledge of the type of
drive. This predictive power is remarkable, since the eigenmode
structure is just an equilibrium property of the passive system.  The
argument based on eigenmodes has been verified in non-polygonal
configurations as well, although we managed to find some exceptions to
this rule~\cite{bruot12b}.  Refs.~\cite{cicuta12,bruot12b} also
present the equations of coupling of the system in terms of
``reduced'' Oseen tensors that take into account that the direction of
each rower is constrained, so that a relation between the $x$ and $y$
motion of a particle allows to reduce the dimension of the coupling
matrix from $2 N \times 2 N$ to $N \times N$.

The Oseen eigenmodes for a given geometrical arrangement are therefore
predictive of the collective motion in the active state: this is a
very powerful result, useful in designing optimal geometrical
arrangements for sustained collective fluid transport by these active
oscillators.  In arrangements of many oscillators, the coupling is
strongest the closer they are geometrically. Therefore in a disordered
and large system it should be possible to identify clusters,
i.e. groups of 2, 3, 4 or 5 spatially close oscillators which are more
tightly coupled together, as in \myfig{fig:models_and_expectations}(b)
(inset).  These will synchronize into ``dynamical motifs'' similar to
the ones we have observed in~\cite{cicuta12z}, perhaps enabling the
larger scale (tissue level) metachronal wave to be understood as
resulting from local units. It will be fascinating in the future to
discover if biology makes use of this hierarchical structure, and to
what level of disorder can be tolerated in a ciliated tissue.  Long
range metachronal waves have also been reported numerically and
analytically in chains of rowers~\cite{bassetti03,stark11}.  It was
shown in~\cite{stark11} that metachronal waves emerged when the range
of the hydrodynamic interaction was short, for example considering
only the interaction between nearest neighbors, as happens in the
limit of being close to a no-slip boundary.


\subsection{Oscillators moving along orbits: The ``rotor'' model}

Another very useful model system is to consider stress-controlled
oscillators.  Here, the key control parameter is force, rather than
position.  This type of model is one step closer to the situation in
biological cilia, where the motors inside cilia also play the role of
sensors for the state of stress on the filament, although perhaps the
``geometric switch'' discontinuity is a good representation of a
collective stress-driven detachment of motors from filaments,
switching from power to recovery stroke.

\subsubsection{The model}

When a cilium beats, its center of drag moves along a given orbit that
can be a 2d or 3d path.  A way to model a cilium is hence to prescribe
an orbit for the particle~\cite{vilfan06,lenz08,golestanian12}.  With
$\phi$ the phase of the oscillator, the orbit is described by its
shape $\mathbf r(\phi)$ and the driving force acting on the particle
$F(\phi)$, that represents the force provided by the cilium, in the
direction tangent to the path at $\phi$, as shown on
\myfig{fig:colloidal_models}(c) for circular orbits.
Additionally, the orbit can be made flexible~\cite{lenz08}, which
helps to break the time-reversal symmetry required for
synchronization.
In order to generate a net flow, the trajectory can be tilted and
positioned close to a surface, so that the drag coefficient changes
during the cycle because of the variable height from the wall. This
idea of the possible importance of wall effects in the flows generated
by cilia and flagella was already emphasized by Blake~\cite{blake82}.

\begin{figure}[t!]
  \includegraphics{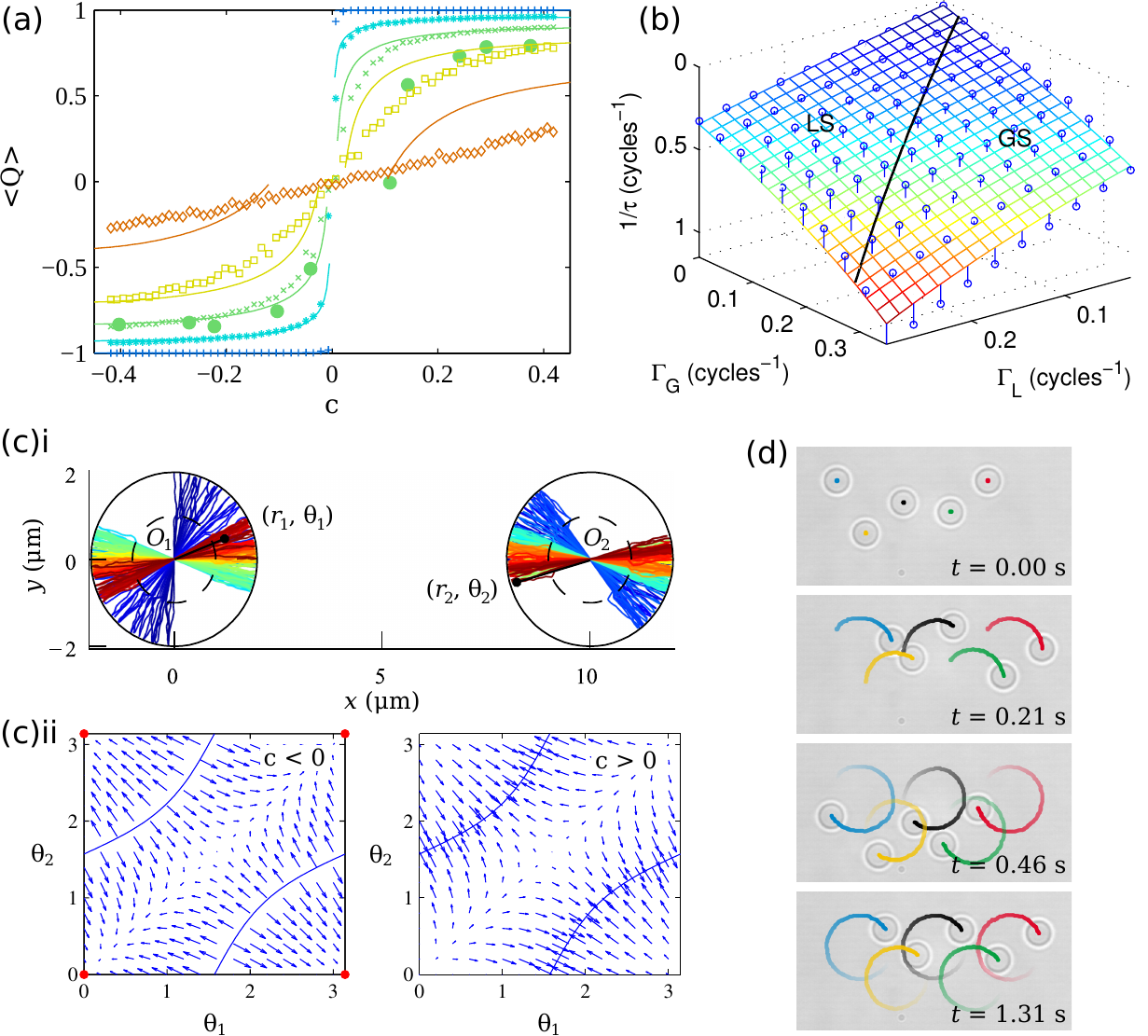}
  \caption{\label{fig:main_results} Main results on colloidal models.
    (a) Synchronization of two rowers.  Depending on the curvature $c$
    of the driving forces, two rowers synchronize in phase or in
    antiphase ($\left< Q \right> = -1$ or 1 respectively) in
    experiments (filled circles), BD simulations (open markers) and
    theory (solid lines).  The different colors indicate different
    temperatures (increasing from blue to red), and higher
    temperatures require a stronger curvature of the driving forces to
    see synchronization.  (b) Synchronization (in phase) of two
    rotors: simulations (markers) and theory (colored surface).  The
    inverse relaxation time of the phase difference $1/\tau$ is a
    measure of the strength of synchronization.  Here, the strength of
    synchronization is controlled by two dimensionless parameters
    $\Gamma_{\text L}$ and $\Gamma_{\text G}$ for the Lenz and
    Golestanian mechanisms, respectively.  The mechanism that
    dominates is indicated by the LS and GS areas (Lenz and
    Golestanian synchronization).  (c) Alignment of two rowers with a
    free direction of beat.  (c)i Tracks of two beads in the case of
    negatively-curved potentials from a random initial condition
    (increasing time from blue to red).  Each oscillator is a rower
    with a free direction of oscillation.  Here, the beads oscillate
    along diameters of a circle and slowly align horizontally,
    i.e. $(\theta_1, \theta_2)$ tends to $(0, 0)$.  (c)ii Analytical
    convergence map of the $(\theta_1, \theta_2)$ orientation of the
    beats depending on the curvature.  For $c < 0$, the convergence to
    $(0, 0)$ (red dots) is predicted, while for $c > 0$, the pair of
    angles converges to a configuration on the solid lines.  In the
    presence of Brownian noise, this system does not align, but
     stays close to  the solid lines.  (d) 5-rotor configuration
    that synchronizes after a few cycles.  To make synchronization
    possible, the driving force on the central rotor needed to be
    reduced, as it is more coupled than others to the rest of the
    system.}
\end{figure}

\subsubsection{Understanding synchronization in rotors: role of force
  profile}

In~\cite{vilfan06}, Vilfan and J\"ulicher showed with numerical
simulations that two beads driven on elliptical and tilted
trajectories near a wall can synchronize, with their state controlled
by the relative position of the orbits.  In this case, synchronization
comes from the hydrodynamic coupling between the two oscillators that
allows the particle to move faster or slower on its trajectory than if
it was just pulled by its driving force.  This way of synchronization
was later addressed by Uchida and Golestanian, who derived generic
conditions for synchronization~\cite{golestanian11a,golestanian12}.
Here, a trajectory is simply defined by $\mathbf r(\phi)$ and
$F(\phi)$ that are both functions of a phase $\phi$ and no flexibility
of the orbit is assumed.  The hydrodynamic coupling is described by a
``geometric factor'' $H_{12}(\phi_1, \phi_2)$ which is the
hydrodynamic tensor projected along the tangents to the two
trajectories at $\phi_1$ and $\phi_2$, where 1,2 designate the two
oscillators.  By linear stability analysis, they obtained that the
condition for synchronization of the two oscillators is that the
growth rate
\begin{equation}
  \Gamma \approx -\frac{2}{T_0} \int_0^{2 \pi} \text d \phi \, \left[
    \ln F(\phi) \right]' H_{12}(\phi, \phi)
\end{equation}
is negative~\cite{golestanian12}, where $T_0$ is the intrinsic period
of the oscillators.  This formula highlights that both the driving
force and $H_{12}(\phi, \phi)$ must depend on $\phi$ to obtain
synchronization.  Since for most trajectories the second condition is
satisfied~\cite{golestanian12}, the condition usually reduces to
having a non-constant $F(\phi)$.  The formula also gives the intuition
that near-field effects can be used to synchronize the system by
having strong variations of the geometric factor.  By pushing the
calculation further, they also defined an effective potential of
synchronization $V(\Phi_1 - \Phi_2)$ that allows to quickly identify
the synchronized states (minima of this potential) and the likeliness
of phase slips in the presence of noise (they depend on the amplitude
of the potential and its possible tilt).  Here $\Phi_i$ is a gauge
linked to $\phi_i$, such that $\Phi_i$ increases linearly if the
oscillator is uncoupled.

In the case of two circular trajectories (as in
\myfig{fig:colloidal_models}(c) but without flexibility) Uchida
and Golestanian showed that in the Fourier decomposition of the force
profile, the mode that leads to strongest synchronization
is~\cite{golestanian12}
\begin{equation}
  \label{eq:optimal_golestanian_profile}
  F(\phi) = F_0 \left[ 1 - A_2 \sin (2 \phi) \right]
\end{equation}
with $0 < A_2 < 1$.  This is valid in the far-field limit, when the
size of the orbits is much smaller than the distance $d$ between the
oscillators, and when the oscillators are either very close to the
surface ($h \ll d$, with $h$ the height from the surface) or in the
bulk ($h \gg d$).  For these circular paths, if $A_2 = 0$, the
amplitude of the driving force is constant and the system does not
synchronize.

More recently, this model was used by Golestanian and Bennett to study
the synchronization of \emph{Chlamydomonas} in simulations: Two
colloidal rotors were maintained close to a third sphere that is
modelling the cell body. In the presence of noise, this lead to a
run-and-tumble behavior~\cite{bennett13}, similar to the actual
alga. The synchronized states and their stability have also been
studied without noise, depending on the choice of the driving force
profile~\cite{bennett13a}.

\subsubsection{Understanding synchronization in rotors: role of
  trajectory compliance}

As just explained, synchronization can arise from particular choices
of the force profiles driving rotors.  However, synchronization in
rotors can also emerge from another mechanism: the flexibility
(compliance) of the orbits.  Flexibility leads to loss of
time-reversal symmetry, and is a common way to force hydrodynamic
synchronization in various systems such as rotating helices and
paddles~\cite{reichert05,reigh12,qian09}.  It has been proposed for
colloidal rotors by Niedermayer, Eckhardt and Lenz in~\cite{lenz08}.

This model is shown in \myfig{fig:colloidal_models}(c), where the orbits are circular and
$F(\phi_i)$ is now set to a constant.  Flexibility is allowed in the
direction orthogonal to the orbit, meaning that instead of following
exactly the circular trajectory, the
particles can deviate from their orbits, for example because of the
coupling with other particles or thermal fluctuations. To stay close to the predefined path, a
restoring force is added which tends to pull the particle back to the
track. The restoring force can, to the lowest order, be described by a
spring constant as shown in \myfig{fig:colloidal_models}(c).  For two
rotors driven by constant forces and without thermal fluctuations,
Niedermayer, Eckhardt and Lenz calculated the state of synchronization
and its strength.  They found that the decay rate of the phase, as it
converges to the synchronized state (in phase), is proportional to the
inverse of the spring constant $k_r$ that constrains the beads to the
circles. In long chains of oscillators, their simulations also led to
metachronal waves in both cases of periodic and free boundary
conditions at the ends of the chain.

\subsubsection{Towards more elaborate rotor models}

The early rotor models represented above were highly simplified as
they either assumed stiff orbits (``Golestanian model'') or constant
driving forces (``Lenz model'').  These idealized views have been very
useful to determine the origin of synchronization.  However, a more
realistic representation of a cilium should include both non-constant
driving forces to represent well the power and recovery strokes of a
cilium, and flexibility.  This has been done experimentally with
optical tweezers in~\cite{bruot13b}.
In \myfig{fig:colloidal_models}(c), for the oscillator $i$, the center
of a trap is maintained ahead of the particle at a distance
$\epsilon(\phi_i)$.  The trap is controlled by a feedback loop that
ensures that it is positioned, for every feedback cycle, at
$\epsilon(\phi_i)$  along the tangent to the preprogrammed orbit, at
the point at $\phi_i$ at which the bead is.  Controlling the trap-to-bead
distance sets the driving force $F(\phi_i) = k \epsilon(\phi_i)$, where
$k$ is the trapping constant of the optical tweezers.
Some degree of flexibility in the trajectory is always present when
using optical tweezers.  If the optical traps have the same stiffness
in all directions of the focal plane, the flexibility would be simply
$k$.  Thus Golestanian and Lenz synchronization can be tuned
independently by changing $k$ and $\epsilon(\phi_i)$.  In practice, we
have implemented complex trapping landscapes by time-sharing a laser
beam with acousto-optic deflectors (see~\cite{cicuta12t} for details
on this method).  This allowed us to study oscillators with much
higher flexibility than what we would have obtained with simple
harmonic traps.
With this setup, we have been able to observe both Lenz and
Golestanian in-phase synchronization in the configuration shown in
\myfig{fig:colloidal_models}(c) with the force profile in
\myeq{eq:optimal_golestanian_profile}.  In an intuitive
analysis, we have successfully recovered the synchronization strength
$\Gamma$ (i.e. the cycle-averaged decay rate of the relative phase
difference at the synchronized state) as a simple sum of the
Golestanian and Lenz strengths of synchronization $\Gamma_G$ and
$\Gamma_L$ of the two models described above:
\begin{equation}
  \Gamma \approx \Gamma_G + \Gamma_L =  2 \pi \frac{3 a}{4 d} \left(
     A_2 + \frac{3 F_0}{k r} \sqrt{1 - A_2^2} \right)\, ,
\end{equation}
with $a$ and $r$ the particle and orbit radii, and $d$ the distance
between the oscillators.  \myfig{fig:main_results}(b) shows a
comparison between the measured inverse relaxation time $1/\tau$ of
the phase difference in simulations (circles) and the expected rate
$1/\tau \approx - \ln [1 - (\Gamma_L + \Gamma_G)]$.

\subsection{Alignment of the beating planes: The ``rower'' with a free
  direction}

Synchronization of  the beads is not the only cooperative behavior that coupled
oscillators can display.  For instance, in a carpet of cilia showing a
metachronal wave, all the cilia roughly beat in the same direction.
The direction of cilia beats in a fully grown epithelial tissue is
well defined relative to the organ.  This is essential for the
generation of fluid flow, for example for mucus clearance away from
the lungs, which relies on coordinated beating of cilia to produce
transport-efficient metachronal waves~\cite{button12}.  However, the
origin of this orientation that develops during the tissue growth
(after planar cell polarity is established) is still an open question
in developmental biology.  For example, the network of microtubules
connecting the basal bodies could couple to the cell
shape~\cite{vladar12}, and the hydrodynamic interaction may also play
a role in setting the orientation~\cite{guirao07,guirao10}.

In the rower view of a cilium each oscillator has a direction of
oscillation.  While this direction was constrained in the work
described in Section~\ref{sec:rowers}, we have also modified the model
to allow the direction to deviate slowly because of hydrodynamic
forces (details of this extended model can be found
in~\cite{bruot13a}).

The rower model with free orientation produces rich results.  We saw
that two oscillators synchronize in phase and align in the direction
defined by the line joining the centers of the oscillators, if they
are driven by power-law potentials with an exponent $\alpha < 1$ ($c <
0$, see \myfig{fig:main_results}(c)i).  The figure shows the tracks of
the two oscillators starting from an initial condition (dark blue)
with a random phase and a random orientation.  After a few tens of
oscillations, the system converged to parallel oscillators.  The
in-phase synchronization (not clearly visible on the figure) is
consistent with results on rowers with fixed orientation, where
in-phase locked states were obtained for negative curvatures $c$
(\myfig{fig:main_results}(a)i).  For $\alpha > 1$ ($c > 0$), the
system does not converge to a fixed orientation: it rather chooses
orientations for which the hydrodynamic coupling vanishes, and as a
consequence synchronization becomes weak.  The convergence of the
orientations are summarised in \myfig{fig:main_results}(c)ii, where
the arrows show the map of the time evolution of the orientations
$\theta_1$ and $\theta_2$ of the two oscillators.
We also simulated configurations of up to 64 oscillators with a free
orientation, and systems with $\alpha < 1$ always aligned strongly,
even with noise, in the direction of the elongation of the array of
oscillators.  This could be relevant to explain the alignment of cilia
in elongated cells such as \emph{Paramecium}, and perhaps in planar cell
polarized tissues, see Section~\ref{sec:outlook}.

\subsection{Limitations of the rower and rotor models}

An important limitation of the colloidal models is that it is hard to
implement experimentally colloidal oscillators  in a way
 that they generate a net fluid flow, since typically a spherical
bead is undergoing a back-and-forth motion. Swimmers or micropumps in
which the traps are switched with a known frequency (hence no free
phase oscillators), were however realized experimentally by
Cicuta~\emph{et al.}~\cite{cicuta09a,cicuta10b} with two or three
beads, based on model swimmers proposed
in~\cite{najafi04,golestanian08,golestanian08a,earl07}.  In an
extensive study, Pande and Smith showed that synchronization between
the arms of a three-bead swimmer can maximize its net
velocity~\cite{pande15}.  A model oscillator that generates a net flow
should have many degrees of freedom, and good candidates for a model
may be self-assembled magnetic colloids~\cite{vilfan10,vilfan12}.

The current challenge is  to develop experimental colloid models
that account for more realistic biological conditions. Some steps
towards that have been made with the rower with a free direction of
oscillation, or with the circular rotor model that includes both a
non-constant driving force and flexibility.  An extension of the rotor
model could investigate the case of non-circular trajectories
experimentally.  The motion of the center of mass of a
\emph{Chlamydomonas} flagellum is indeed far from circular (see
supplementary material of~\cite{cicuta12t}).
More importantly, the effect of the surface on which the cilia are
anchored is often neglected.  The main effect of the surface is to
shorten the range of the hydrodynamic coupling, from a $1/d$ to a
$1/d^3$ decay between a bulk fluid and close to the surface.  While
this is not very relevant when studying only two oscillators if the
typical size of the oscillations is small compared to the distance
between oscillators, as it only changes the amplitude of the coupling,
it can have an effect on the collective dynamics of more than two
oscillators~\cite{stark11}.
Finally, many biological fluids that surround the cilia, like mucus,
are highly non-Newtonian, and in some cases not even homogeneous.
This can affect both the motion of a single cilium and the collective
properties of an assembly of oscillators.  While experiments similar
to those described above could be simply reproduced in a viscoelastic
fluid, simulating the model and obtaining analytical determinations of
the synchronization properties will pose entirely new challenges.

\section{BIOLOGICAL SYSTEMS AND LINKS WITH MODELS}
\label{sec:sync_in_real_systems}

Particular strategies have been developed by microorganisms to enable
motility in liquids. The most well studied cases are bacteria
flagella~\cite{darnton04} and the eukaryotic cell cilia, reviewed
in~\cite{lauga09}.  Few experiments exist on waves involving large
numbers of real cilia, with resolution at or close to individual
units: Okamoto and Nakaoka reconstituted cortical sheets of
\emph{Paramecium} ciliated cells, and observed the emergence of
collective cilia beating~\cite{okamoto94a}, Berg and coworkers have
studied ``bacteria carpets''~\cite{darnton04}. The airway tissue is an
``intrinsic'' and bio/medically relevant system and is experimentally
accessible. This section aims to overview the experimental
observations that have been made on increasingly complex (larger cilia
number) living systems, and the models that have been proposed to
explain synchronized collective motions.

\subsection{Single cell organisms with few cilia}

With only two flagella, \emph{Chlamydomonas} has been a successful
system to investigate synchronization.  This organism can be easily
handled experimentally, and many mutants are available to tune several
parameters, like making it uniflagellar~\cite{vincensini11}.  Relevant
information can be obtained both to gain access to the swimming cycle
of the alga and to characterize the synchronization of the flagella.
For example, we have used data on the cycle of beat of a
flagellum~\cite{bayly10} to determine the shape of the driving
potential that can be inputted in the rower model.  Using this
potential lead to a prediction on the synchronization state consistent
with the breaststroke motion of \emph{Chlamydomonas} that dominates in
the wild type of the alga~\cite{cicuta12t},
see~\myfig{fig:cilia_overview}(b).
Goldstein \emph{et al.} followed another approach --- characterizing
the synchronization.  They observed that the swimming of the alga also
includes short periods of phase drifts during which the flagella
oscillate asynchronously~\cite{goldstein09a,goldstein09b}.  This is
linked to the organism's motility: \emph{Chlamydomonas} displays
periods of long straight trajectories, alternating with abrupt changes
of direction, similarly to the ``run-and-tumble'' motility well known
in prokaryotes~\cite{goldstein09a}. A recent study found consistency
between this behavior of the two flagella and a generic model of two
coupled noisy phase oscillators with a coupling strength consistent
with hydrodynamic interactions~\cite{goldstein09b}, and the tumble
looks similar to the antiphase state seen in the \emph{ptx1}
mutant~\cite{goldstein14c}.  Hydrodynamic forces are certainly
sufficient to induce synchronization of the flagella as beautifully
demonstrated in~\cite{goldstein14}, but there is also evidence that in
the swimming state the mechanical coupling through the motion of the
central cell body also needs to be considered~\cite{friedrich12}.

\subsection{Metachronal waves in \emph{Paramecium} and \emph{Volvox}}

\emph{Paramecium} and \emph{Volvox carteri} are respectively
unicellular and multicellular organisms, a few hundreds of micrometers
in size, with an outer surface covered with thousands of
cilia~\cite{vincensini11,kirk98}: They are model organisms for imaging
metachronal waves.

In the colonial alga \emph{Volvox}, each of the ciliated cells
composing the carpet of filaments is biflagellated. These two
organisms show an interesting collective motion, in which the flagella
do not all beat with the same phase. Instead, two neighboring flagella
will keep a constant, usually small, phase difference. When
oscillating, this leads to the formation of a metachronal wave that
can be seen at the surface of various cells or
tissues~\cite{blake74a,tamm75,okamoto94a,gueron97,aiello72,sanderson81}.
In \emph{Volvox}, this propagating phase pattern is also perturbed by
defects (local phase shifts that appear and disappear periodically)
and this dynamics has been reproduced in a system of hydrodynamically
coupled tilted rotors near a surface~\cite{brumley15}.

In-phase synchronization and metachronal waves are believed to
optimize the flow generated by an assembly of
cilia~\cite{osterman11,mettot11}.

\subsection{Mammalian mucociliary tissues}

In mammalian cilia (in the brain ventricles, fallopian tubes, airways,
etc.), the cilia belong to multiciliated cells, and there are
typically of order 200 cilia per cell, and each cilium is separated by
approximately 200\,nm. These ciliated cells, together with
mucus-producing cells (goblet cells, in mammals), and ion-regulating
cells, form a general tissue type known as mucociliary
epithelium. Experiments have considered developing embryos of
mice~\cite{francis09,stubbs12}, and closely related phenomena take
place on the outer surface of various ``model'' biological organisms,
such as the \emph{Paramecium}~\cite{beisson99} and the algae colony
\emph{Volvox}~\cite{goldstein15c} as well as and \emph{Xenopus}
frogs~\cite{dubaissi11}.

\begin{figure}[t!]
  \includegraphics{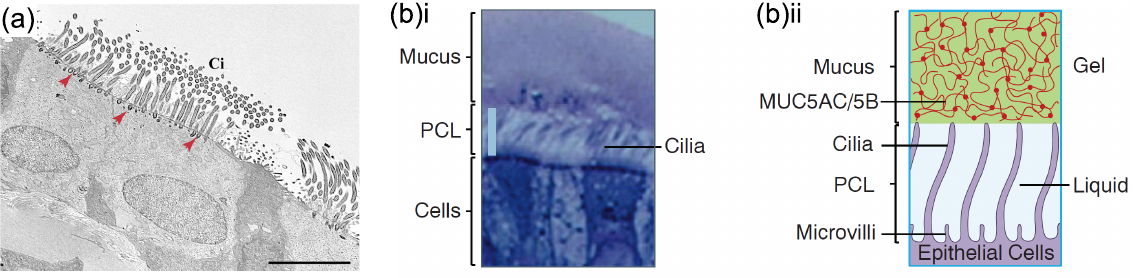}
  \caption{\label{fig:cilia_side_view} Side views of trachea
    epithelial cilia.  (a)~Mature ciliated cells in rat
    trachea~\cite{dawe07}.  During ciliogenesis, centrioles develop
    into basal bodies (red arrows).  These have extended cilia (scale
    bar: 5\,$\mu$m).  (b)~Periciliary layer (PCL)~\cite{button12}.
    The gel-like mucus phase ``sits'' on the cilia brush without
    extending in the cilia layer.  Instead of mucus, the cilia are
    surrounded by a liquid phase, the PCL (scale bar: 7\,$\mu$m).}
\end{figure}

Apart from understanding their synchronization and cooperative
behavior, ciliated tissues raise other interesting questions: for
example, in multiciliated cell tissues (carpets of many cilia) the
cilia all beat in parallel planes, posing the question of how the
direction is chosen in tissue development~\cite{vladar12} and
maintenance.  Cilia grow outwards from a structure called the basal
body, which is anchored to a cell's cytoskeleton~\cite{mizuno12}
(\myfig{fig:cilia_side_view}(a)). The basal body is itself generated
from centrioles, which are subcellular structures responsible for
microtubular organisation.  Once fully grown, a cilium transverses the
cell's plasma membrane extending typically several microns out of the
cell body.
In fully developed epithelial tissue, the direction of beating is well
defined relative to the organ, for example it is parallel to the
trachea in mammalian airways. This is essential for the generation of
fluid flow (mucus clearance away from the lungs in this case) which
relies on coordinated beating of cilia to produce transport-efficient
metachronal waves~\cite{button12}, similarly to \emph{Paramecium}
and \emph{Volvox}.

\subsection{Respiratory tract}

Particularly important in terms of human health, and impact of related
disease, is the respiratory tract. The mucociliary clearance process
is the first line of defence against a wide range of potentially
harmful agents including bacteria, viruses, and particulate
matter. The layer of mucus (ions, water, mucin proteins), kept in
motion by beating cilia and occasionally by cough, acts as a physical
barrier. Flow of fluid is maintained over all the airway surfaces,
toward the larynx. As reviewed in~\cite{livraghi07}, dysfunction in
this clearing system, which can arise from conditions such as human
cystic fibrosis (CF) and primary ciliary dyskinesia (PCD), has severe
consequences such as chronic lung infections and respiratory
insufficiency. Altered mucus clearance is also linked to other chronic
airway diseases such as asthma that afflict millions worldwide.

\subsubsection{Mucus clearance by cilia}

The human bronchial system is complex: epithelium is mucociliated and
pseudo-stratified \emph{in vivo}, and the mechanisms that regulate its
functions are not fully understood. Among the epithelial cells, goblet
cells secrete mucus, which protects the bronchial epithelium and is
transported by the motile cilia protruding from the apical poles of
specialized epithelial cells (\myfig{fig:cilia_side_view}(a)). The
coordinated system of epithelial cells --- including functions as ion
transport through their membranes, mucus secretions and cilia action
--- and cough, is collectively termed mucus clearance, and results in
a continuous flow. The respiratory tract is lined by a thin layer of
airway surface liquid, which is approximately 7-70\,$\mu$m in
height~\cite{button12} (\myfig{fig:cilia_side_view}(b)).
This layer is itself defined as consisting of two phases: an outer
mucus-rich layer, and an inner fluid close to the cilia, known as
``periciliary liquid layer'' (PCL). The PCL height is in practice
defined as height of the extended cilium, and recent studies have
demonstrated complex interactions among cilia, the mobile mucins in
the mucus layer, and the underlying
PCL~\cite{thornton08}.  It
appears that both the mucus layer and the PCL are moved
unidirectionally, and it is suggested that the movement of
the
airway surface liquid
(ASL)  involves two steps~\cite{button12}: First, the ciliary
power stroke acts on the under-surface of the mucus to move the mucus
layer unidirectionally on the airway surface; second, the frictional
interaction of the mucus layer with the PCL allows this underlying
layer to travel at velocities similar to those of the overlying mucus
layer (a form of ``secondary'' transport). These data are consistent
with the fact that mammalian airway cilia are too short, and their
recovery stroke not sufficiently close to the cell surface, to move
the PCL directly.  The flow within the boundary PCL layer itself is
very complex, and has been addressed from a colloidal model
perspective by~\cite{vilfan10}.

\subsubsection{Analysis of collective motion on airway cells}

Various types of experiments are possible relating to airway cells:
(a)~sections of the trachea can be taken from animals, and maintained
in a living functioning state for over a day; (b)~human cells can be
obtained by non-invasive procedures (nose brush or scrape) and
cultured to recreate an \emph{in vitro} tissue; (c)~ciliated cell
lines, and fully confluent tissues, are available commercially.  Cells
in all these \emph{in vitro} systems express their normal mucus layer
(which can be temporarily washed-off by buffer).  The choice of
experimental approach depends on the question of interest.

The physical properties of mucus, including rheological parameters,
are important in setting the frequency of cilia beating
and they have been investigated extensively in the context of various
pathologies~\cite{tilley15}. For example, in cystic fibrosis there is
clear evidence that the properties of the PCL are abnormal.
Existing observations in the literature of ciliated tissues have
focused mainly on measuring cilia beat frequency and mean flow
velocity.

\section{OUTLOOK}\label{sec:outlook}

A key objective is to link quantitatively the emergent collective
dynamics to the behavior of the individual oscillators, and to the
physical properties of the materials surrounding the cilia.  Current
models are minimal, and realistic aspects of the real systems will
need to be examined and included into the
models, such as
viscoelasticity of the liquid, the presence of nearby
surfaces~\cite{vilfan10}, heterogeneity of cilia
distribution~\cite{golestanian10b}, plasticity of beat
orientation.
 Linking all the scales of this problem, i.e.  how collective dynamical phenotypes emerge from the
molecular functioning of the cilium,  remains  a formidable
long-term challenge, requiring  combined experiments on
colloidal models, simple biological organisms, and complex biological
tissues. Understanding this link will be a breakthrough in the physics of
synchronization and will open up diagnostic potential
 for example in the context of pathologies
 of the airway tract.




\subsection{Clinical diagnosis of diseases}
{Clinical diagnosis most commonly consists of inspection of cilia beat frequency, and cilia waveform,
from small biopsies. In cases of suspected genetic disease (PCD), the ciliar molecular structure is assessed
by electron microscopy.
Key aspects of cilia dynamics (such as space and time coherence)  are currently not being addressed in
the clinical setting, but they  are evidently important
 in conditions leading to  ``patchy loss'' of cilia, which can be a consequence of
explosion trauma or infections. They will  also be key in understanding the efficacy of drugs used to treat symptoms of
 CF or asthma, and in future regenerative medicine to
aid development  of  functional tissues and define standards.     } The role of
fundamental physics in this context is to highlight which dynamical
patterns and which parameters of the driving force are important, so
that these significant parameters can be investigated and measured in
the biological system.

 One challenge is to go beyond this analysis,
and to extract more information, which would feed into a two-way link
between the collective steady state and the function at the single
cilium level. In this way, from easily measurable macroscopic
behaviors, one would be able to infer microscopic properties. For
example, recent developments push towards resolving the shape of a
single cilium during the power and recovery strokes, which is now
possible in simple organisms~\cite{bayly10}. These experiments are
important in relating to the microscopic molecular function within a
cilium. If one had high resolution observations of the shape of a
beating cilium during each stroke, and knowledge of the mechanical
properties of the filament bundle, this could be linked quantitatively
with models that account for cilia structure and molecular motor
activity such as~\cite{hilfinger08}. However another question remains,
which is at least as important because it too offers a (possibly
experimentally easier) way to link the macroscopic to the molecular
scale: given a force versus time profile throughout the stroke (which
can be thought as an intrinsic property of a certain cilium in given
external physical conditions), what is the resulting collective
property of the tissue surface, in which many such cilia are spatially
arranged?

Answering this challenge requires bringing together fundamental ideas
and carrying out clear experiments on the biological systems to prove,
control and demonstrate the relevance of what is identified from the
colloid model experiments. A clear understanding of how the bottom-up
emergence of certain dynamical motifs (from given force profiles,
geometrical arrangements, and coupling forces) is achieved, will
provide a powerful tool: a map that makes it possible to relate these
motifs back to those microscopic properties, which could then feed into a video analysis
package that could be deployed in the clinical setting for the study
of pathologies linked to ciliary dysfunction.

\subsection{Development of ciliated tissues}

An open question in developmental biology is to find the rules and the
cues that enable cilia tissues, a fairly complex, and well organized
carpet, to be made, and in particular, how the direction of the cilia
is chosen.
The first symmetry to be broken in the development of vertebrates is
the anterior/posterior. For example the planar cell polarity (PCP)
pathway sets the initial direction in a developmental stage of the
epithelium in \emph{Xenopus} embryos, a tissue which includes
multiciliated cells~\cite{mitchell09}.  From this point on, there are
gradients of a variety of biochemical elements along this axis. Cells
can be polarized, both in the intracellular protein localization, and
in their shape. This process happens before cilia-genesis, and the
standard view in biology is that the gradients in biochemical markers
control most, if not all, of the subsequent organ
development. However, once cilia are generated, they contribute to
long-range flows, which can transport chemical factors directionally,
or act as a mechanical cue for
organization~\cite{mitchell07,marshall08,wallingford10,freund12}.

In the specific case of developing orientational cilia order in the
airway tissue, there is a hypothesis that flow-induced self
organisation might be important.  A fraction of the cells in the
tissue that will develop into the airway epithelium express a few
hundred centrioles, which become basal bodies and grow cilia.  At this
stage, the cells themselves are already polarized (biochemically and
in shape) but the basal bodies when they first appear are not
oriented. There have been very recent studies suggesting that the
network of microtubules connecting the basal bodies could couple to
the cell shape or to the emerging tissue architecture, and possibly
orient the cilia~\cite{vladar12}.  On the other hand, the newly made
cilia are exposed to a directional flow from the mucus being produced
by other cells. These cilia will also be exposed to the flow that they
themselves generate, i.e. they interact with each other through
hydrodynamic interaction forces~\cite{gueron97}.  The question of how
are cilia aligned has also been addressed looking at mouse brain
ventricles~\cite{guirao10}, where it was shown that cilia first dock
apically with random orientations, and then reorient in a common
direction through a coupling between hydrodynamic forces and the PCP
protein Vangl2.

The evidence of the role of flows in determining the orientation of
cilia is therefore present in the experiments in the \emph{Xenopus}
larval skin~\cite{mitchell07,werner12} and in mouse brain
ventricles~\cite{guirao10}. However, there are only very few physical
models to explain this behavior~\cite{guirao07}.
Experiments where  air or liquids are driven across the cell
laden surfaces  will provide insight  as to whether and at what development stages
these stresses can
reorient cilia attachment and resulting active transport.

\subsection{Basic science and microfluidic technology}

The models discussed here, showing synchronization regimes that are
tunable, and competing with Brownian noise, are also very accessible
platforms to develop the science of nonlinear
systems~\cite{pikovsky01}. In this, they have the ``usual'' great
beauty in common with many colloidal particle model systems: Easily
accessible timescales, direct imaging of configurations, intrinsic
presence of stochastic elements (through thermal noise). Our
understanding of which configurations, geometries and coupling forms
might be ideal to maximize susceptibility, or create active media that
support particular signal propagations, is still very crude.

In the technological arena, taming appropriately the mechanisms of
hydrodynamic synchronization could make it possible to optimize    devices
such as  micro-pumps~\cite{onck11}, molecular sorters and  low-Reynolds
number mixers, and control behavior of swarms of swimmers.  Various
groups have succeeded in generating artificial carpets of beating
filaments~\cite{masegosa14}, but not yet with free phase.  If such
systems had spontaneous synchronization, this might be designed in
such a way that it can sustain fluid drive over a variety of
conditions (as do the biological mucociliary tissues), whereas a
deterministically driven carpet is necessarily tuned for just one
condition.

\begin{summary}[SUMMARY POINTS]
  \begin{enumerate}
  \item Model systems show that hydrodynamic forces, exerted by one
    cilium on another, can be sufficient to cause synchronization, and
    emergence of collective wave states.
  \item The properties of the emergent collective behavior depend on a
    number of factors, which can be individually addressed
    (experimentally, numerically and analytically) working with simple
    models, in which each motile cilium is coarse grained into the
    movement of a single colloidal sphere, retaining elements of the
    original system in the details of the drive and geometric
    disposition.
  \item Two main classes of models (rowers and rotors) have been
    studied so far; both can be mapped onto biological data, both have
    rich phenomenology, and each has particular aspects that make the
    model attractive.
  \item Model systems have been used to highlight the regime of
    synchronization, which is limited by thermal noise, and by
    detuning of the natural frequencies of the driven oscillators.
  \item Model systems are also able to explain the type of
    synchronization; the main thrust of work so far has revolved
    around mapping the system of the biflagellated algae
    \emph{Chlamydomonas}, and addressing how it manages to exhibit
    both in-phase and antiphase beating patterns.
  \end{enumerate}
\end{summary}

\begin{issues}[FUTURE ISSUES]
\begin{enumerate}
\item Our fundamental understanding of this class of non-linear
  systems is still incomplete, and the aim of linking local rules to
  the macroscopic properties of emergent dynamics is in general still
  an open challenge.
\item The class of non-linear systems described here has so far been
  explored as models of biological motile cilia, but there are many
  other future areas of development in technology (artificial surfaces
  that drive fluid flow) and micro-scale sensing (optimized arrays for
  picking up faint signals).
\item Understanding robustness against geometric heterogeneity: In
  ciliated tissues, there are clusters of nearby cilia (e.g. from the
  same cell), and longer distances in between ciliated cells; in some
  diseases, patchy loss of cilia leads to regions devoid of beating
  cilia.
\item Hoping to describe ciliated tissues such as the airways, it will
  be necessary to understand the role not just of mucus
  viscoelasticity, but also of its complex stratified nature.
\item This area of work is partially justified by clinical challenges,
  and aims to improve diagnostics and perhaps even therapeutic
  practice relating to conditions where mucociliary clearance is
  compromised.
\end{enumerate}
\end{issues}

\section*{DISCLOSURE STATEMENT}

The authors are not aware of any affiliations, memberships, funding,
or financial holdings that might be perceived as affecting the
objectivity of this review.

\section*{ACKNOWLEDGMENTS}

We thank all of our collaborators working in this area, particularly
M.~Cosentino Lagomarsino for introducing us to this question many
years ago, and many discussions since then, and L.~Feriani and
M.~Chioccioli for providing the images in
\myfig[s]{fig:cilia_overview}(c)i and
\ref{fig:models_and_expectations}(b).  This work is funded by the ERC
Consolidator Grant HydroSync.

%
%

\bibliographystyle{ar-style1}
\bibliography{BIBDATAv50}

\end{document}